\documentclass[12pt]{article}
\usepackage[latin1]{inputenc}
\newcommand{\text}{\rm}
\begin{document}
\title{How to get rid of Dirac worldsheets in the Cho-Faddeev-Niemi representation of $SU(2)$ Yang-Mills theory}
\author{A.~L.~L.~de~Lemos,~M.~Moriconi,~L.~E.~Oxman\\ \\Instituto de F\'{\i}sica, Universidade Federal Fluminense,\\
Campus da Praia Vermelha, Niter\'oi, 24210-340, RJ, Brazil.}
\date{\today}

\maketitle
\begin{abstract}
In this paper, we present an exact procedure to deal with Dirac strings or worldsheets in gauge theories containing ensembles of mono\-poles interacting with charged fields. For $SU(2)$ Yang-Mills theory, initially we construct the appropriate change of variables of the charged fields (including charged ghosts and auxiliary fields) so that the only change in the integrand of the partition function, in the Maximal Abelian gauge, is the addition of given closed Dirac worldsheets. Next, we derive our main result, namely, we show that it is always possible to choose them in such a manner that the total (open plus closed) Dirac worldsheets explicitly decouple from the charged sector, leaving only the effect of their associated gauge invariant borders (where the monopoles are placed), without missing any information about the center vortex sector. 

This procedure serves as a simplifying basis to deal with ensembles of monopoles and center vortices in the framework of the Cho-Faddeev-Niemi gauge field decomposition, by writing the partition function only in terms of the physical part of the defects to be integrated.
\end{abstract}

\noindent
{\bf keywords}: Nonabelian gauge theories; monopoles and center vortices; Cho-Faddeev-Niemi decomposition.

\section{Introduction}

Every now and then we are faced with field theories containing a charged sector interacting with monopole-like defects. The most remarkable example is associated with the scenario of dual superconductivity for confinement in $SU(N)$ Yang-Mills theories \cite{N}-\cite{hooft1}, \cite{cp}-\cite{KLSW}.

In these theories, the charged sector corresponds to the ``off-diagonal" modes living in the Cartan subalgebra of the nonabelian group, while mono\-poles arise as defects when defining abelian projection gauge fixing conditions \cite{ap}. Monopoles can also be introduced as defects of the local color frame $\hat{n}_a$, $a=1,2,3$, to decompose the gauge fields \cite{cho-a,FN}, \cite{cho2}-\cite{Shaba}. This procedure has the advantage of not relying on any a priori gauge fixing condition. 

In both situations, we have to deal with the associated Dirac strings or worldsheets, depending on whether the monopole defects are point-like or loop-like. Considering that these objects are not observable (their location can be changed by means of a topologically trivial gauge transformation) a natural question that arises is about the possibility of representing physical quantities, such as the partition function, only in terms of their gauge invariant borders (where monopoles are located). 

In this paper, we will present an exact procedure to achieve this goal in compact $QED(3)$ with charged fields and in the framework of the Cho-Faddeev-Niemi decomposition of pure $SU(2)$ Yang-Mills theory. 

This is particularly relevant in the latter case, when using the gauge field decomposition to guide the obtention of effective theories associated with ensembles of monopoles and center vortices (see ref. \cite{LEO}).
In particular, if our procedure were not applied before any approximation scheme, it would be possible that the effective theories obained could make no sense physically, as Dirac worldsheets would become observable because of the approximations.

On the other hand, once we have a partition function representation only in terms of the monopole locations, assuming a  phase where monopoles condense, we can reobtain the effective model of ref. \cite{cho-a}, proposed by following physical heuristic arguments to deal with the Dirac worldsheets. 

Another closely related example occurs in refs. \cite{cho1, kondo.sky}, where the effective Skyrme model \cite{FN, Shaba, F} has been discussed in the Cho-Faddeev-Niemi framework, by following heuristic arguments assuming a magnetic condensate, and by implementing a series of approximations to compute the one-loop effective action in a monopole background. 

In fact, in these references the singular terms where the worldsheets are concentrated are missing (see the discussion in ref. \cite{LEO}). Of course, any heuristic reasoning only deals with physical objects and the effective theory must be directly constructed in terms of them. Then, the effective models have been constructed in terms of the third component $\hat{n}=\hat{n}_3$ of the local color frame, as monopoles can be seen as defects of this component, with no reference to any Dirac worldsheet. 

The main point is that, as discussed in ref. \cite{LEO}, when  monopole defects are present for $\hat{n}$, necessarily the components $\hat{n}_1$, $\hat{n}_2$ must also contain defects, and therefore we have two possibilities: Firstly, we could have Dirac worldsheet defects where the components $\hat{n}_1$, $\hat{n}_2$ rotate twice, as we go close and around them. This corresponds to a magnetic flux $4\pi/g$ carried by the Dirac worldsheet, matching the magnetic flux $4\pi/g$ emanating from monopoles in nonabelian theories. Secondly, it is also possible to attach  monopoles with a pair of center vortices carrying flux $2\pi/g$, which are also given by defects in the components $\hat{n}_1$, $\hat{n}_2$; in this case, when we go around the vortex they rotate once.

Therefore, when looking for effective models written only in terms of $\hat{n}$, if on the one hand no information about unphysical Dirac worldsheets is introduced, on the other, we miss information about the $\hat{n}_1$, $\hat{n}_2$ sector, which contains physical information about center vortex ensembles. For this reason, it is important to have a careful discussion about how to get rid of Dirac worldsheets in the Cho-Faddeev-Niemi decomposition framework, and to understand why this procedure fails to get rid of center vortices, so that they can be associated with interesting phases displaying confinement, $N$-ality \cite{debbio3}-\cite{quandt} or Abelian dominance \cite{LEO}.
Moreover, the interest in looking for possible extensions to the Skyrme effective model is also supported by recent  limitations of this model observed in the lattice \cite{DHW}.

Technically, the above question about the possibility of representing the partition function with no reference to Dirac strings or worldsheets is a nontrivial one, as in a field theory problem the charge current is distributed on the whole Euclidean spacetime. This is in contrast with the problem of representing the path integral for the propagation of a one-particle system, where the relevant electric current is concentrated on the integration path and the Dirac string does not appear, as long as the Dirac quantization condition is imposed.

Initially, we will perform a change of variables with trivial Jacobian that only introduces given closed Dirac strings or worldsheets in the partition function. In compact $QED(3)$ and $SU(2)$ Yang-Mills theory, this will be possible by working in the Lorentz and the Maximal Abelian gauges, respectively, and considering a gauge transformation with multivalued phase $\chi$, satisfying the Laplace equation $\partial_\mu \partial_\mu \chi =0$. The explicit form of this transformation is obtained by means of the expressions obtained in refs. \cite{engelhardt1,reinhardt} to describe closed thin center vortices. In this respect, note that as is well known, the MAG condition, as well as the Landau condition, do not fix the gauge completely. In \ref{YM-sub}, we will discuss this issue in the context of Gribov ideas for the implementation of a properly defined path integral (see refs. \cite{G,UERJ} and references therein).

Next, we will show that it is always possible to choose the closed Dirac strings or worldsheets, in such a way that the total effect is the decoupling of open plus closed Dirac defects from the charged sector, in the integrand of the partition function, leaving only the effect of their associated gauge invariant borders, where the physical monopoles are placed.

This article is organized as follows. In section \S \ref{c}, we review monopoles in compact $QED(3)$ with charged matter and the Cho-Faddeev-Niemi scenario for $SU(2)$ Yang-Mills theory. Section \S \ref{b} is devoted to discuss the associated partition functions in minimal coupling form and to define the gauge fixing conditions. In \S\ref{d},
we separate, by means of a Hodge decomposition, the terms coupling the Dirac strings or worldsheets to the charged sector from those coupling their borders, where the physical monopoles are placed. In section \S\ref{in}, we carefully discuss the Dirac string or worldsheet independence of the partition functions, and show the central result of this work, namely, how to get rid of Dirac defects by decoupling them from the charged sector. Finally, in section \S\ref{conc} we present our conclusions and discuss exactly where our procedure fails to get rid of physical center vortices.

\section{Charged fields and monopole-like defects}
\label{c}

\subsection{Compact $QED(3)$}

As shown in \cite{polya}, pure compact $QED(3)$ is a confining model. Here, we consider its coupling to a charged matter
sector. In this case, the action \footnote{Throughout this paper we work in Euclidean spacetime.} for an instanton/anti-instanton pair is given by,
\begin{equation}
S = \int d^{3}x \Biggl(\bar{D}_{\mu}\bar{\Phi}D_{\mu}\Phi +
\frac{1}{2}(f_{\mu} + h_{\mu})^{2} \Biggr),
\label{action}
\end{equation}
where
\begin{equation}
D_{\,\mu} = \partial_{\,\mu} - iq\biggl(A_{\,\mu} +
C_{\,\mu}\biggr)
\label{1a}
\makebox[.5in]{,}
f_{\mu} = \epsilon_{\mu\nu\rho}\partial_{\nu}A_{\rho}.
\label{6}
\end{equation}
The field $h_\mu$ added to the dual field strength tensor $f_\mu$ in the action (\ref{action}) is such that
\begin{equation}
\partial_{\mu} h_{\mu}= g_m [\delta^{(3)}(x - x^{+}) - \delta^{(3)}(x - x^{-})] ,
\label{1}
\end{equation}
and the vector potential $C_{\mu}$, satisfying
$h_{\mu}=\epsilon_{\mu\nu\rho}\partial_{\nu}C_{\rho}$, can be
introduced only outside a region containing a Dirac string
$x_{s}(\sigma)$, $\sigma \in [0,1]$, running from
$x^{-}$ to $x^{+}$,
\begin{equation}
 x_{s}(0)=x^{-}
 \makebox[.5in]{,}
 x_{s}(1)=x^+.
\label{2}
\end{equation}
In order to extend the vector potential to the whole space $\mathcal{R}^3$, $h_{\mu}$ and $\epsilon_{\mu\nu\rho}\partial_{\nu}C_{\rho}$ must differ by a singular term $d_{\mu}$,
\begin{equation}
h_{\mu}= \epsilon_{\mu\nu\rho}\partial_{\nu}C_{\rho}+ d_{\mu},
\makebox[.5in]{,}
d_{\mu}=g_m \int_{[x_s]} dy_{\mu}\, \delta^{(3)} (x - y),
\label{3}
\end{equation}
This implies that the flux of $d_{\mu}$
through a surface crossed by the Dirac string is $\pm g_m$ .

The independence of physical quantities on the choice of $C_{\mu}$
must also include the independence on the possible associated
Dirac strings. As is well known, this nonobservability implies the
famous Dirac charge quantization condition
\begin{equation}
q = n e
\makebox[.5in]{,}
e= 2\pi/g_m
\label{dirac_condition}
\end{equation}
where $n$ is an integer.

\subsection{$SU(2)$ Yang-Mills and the Cho-Faddeev-Niemi decomposition}

In $SU(2)$ Yang-Mills theory in four dimensions the action is given by,
\begin{equation}
{S_{YM}}=\frac{1}{2}\int d^4 x\; tr\, (F_{\mu \nu} F_{\mu \nu })
\makebox[.5in]{,}
F_{\mu \nu}=F_{\mu \nu}^{a}T^{a}.
\end{equation}
The generators can be realized as $T^a=\tau^a/2$, $a=1,2,3$, where $\tau^a$ are the Pauli matrices, and the field strength tensor is written in terms of the gauge fields $A_{\mu }^{a}$, $a=1, 2, 3$,
\begin{equation}
\vec{F}_{\mu \nu}=\partial_\mu \vec{A}_\nu -\partial_\nu \vec{A}_\mu +g \vec{A}_\mu\times \vec{A}_\nu,
\makebox[.5in]{,}
\vec{A}_\mu=A_\mu^a\, \hat{e}_a
\makebox[.5in]{,}
\vec{F}_{\mu \nu}=F_{\mu \nu}^a\, \hat{e}_a,
\end{equation}
where $\hat{e}_a$ is the canonical basis in color space.

The Cho-Faddeev-Niemi decomposition \cite{cho-a,FN} is done in terms of a general local frame in color space, $\hat{n}_a$, $a=1,2,3$, which can be parametrized by means of an orthogonal local transformation $R\in SO(3)$ ,
\begin{equation}
\hat{n}_a=R\, \hat{e}_a.
\end{equation}
This frame can be used to represent the gauge field $\vec{A}_\mu$ as,
\begin{equation}
\vec{A}_\mu=A_\mu \hat{n}-\frac{1}{g} \hat{n}\times \partial_\mu \hat{n} + \vec{X}_\mu
\makebox[.5in]{,}
\hat{n}.\vec{X}_\mu=0 ,
\label{dec}
\end{equation}
\begin{equation}
\hat{n}_a.\hat{n}_b=\delta_{ab}
\makebox[.5in]{,}
a,b=1,2,3
\makebox[.5in]{,}
\hat{n}\equiv \hat{n}_3,
\end{equation}
where $\vec{X}_\mu$ transforms in the adjoint representation.

The field strength tensor, for the decomposition (\ref{dec}) defined in the whole Euclidean spacetime, is given by, 
\begin{equation}
\vec{F}_{\mu \nu}=(F_{\mu \nu}+H_{\mu \nu}+K_{\mu \nu}) \hat{n}+\vec{G}_{\mu \nu}+\vec{L}_{\mu \nu},
\label{FHK}
\end{equation}
\begin{equation}
F_{\mu \nu}=\partial_\mu A_\nu -\partial_\nu A_\mu
\makebox[.5in]{,}
H_{\mu \nu}=-\frac{1}{g} \hat{n}.(\partial_\mu \hat{n} \times \partial_\nu \hat{n}),
\label{HK}
\end{equation}
\begin{equation}
K_{\mu \nu}=-i g (\bar{\Phi}_\mu \Phi_\nu-\Phi_\mu \bar{\Phi}_\nu)
\makebox[.5in]{,}
\vec{G}_{\mu \nu}=G^1_{\mu \nu} \hat{n}_1 +G^2_{\mu \nu} \hat{n}_2
\end{equation}
with,
\begin{equation}
\Phi_\mu=\frac{1}{\sqrt{2}}(X^1_\mu+iX^2_\mu)
\makebox[.5in]{,}
G_{\mu \nu}=\frac{1}{\sqrt{2}}(G^1_{\mu \nu}+iG^2_{\mu \nu}),
\label{Fi}
\end{equation}
\begin{equation}
G_{\mu \nu}=
[\partial_\mu+ig(A_\mu+C^{(n)}_\mu)]\Phi_\nu -[\partial_\nu+ig(A_\nu+C^{(n)}_\nu)]\Phi_\mu ,
\end{equation}
and the monopole vector potential is given by, 
\begin{equation}
C^{(n)}_\mu=-\frac{1}{g} \hat{n}_1.\partial_\mu \hat{n}_2.
\label{Cmu}
\end{equation}

Finally, $\vec{L}_{\mu \nu}=	-(1/g) \hat{n}\times [\partial_\mu,\partial_\nu] \hat{n}$
is a term concentrated on the defects of the color direction $\hat{n}$ (see ref. \cite{LEO}). In addition, while in refs. \cite{cho2}-\cite{cho5}, $H_{\mu \nu}$ is computed to be $\partial_\mu C_\nu -\partial_\nu C_\mu$, obtaining simpler ``abelianized'' expressions for the field strength tensor, when dealing with gauge fields containing defects this relationship must be revised. In fact, when defined on the whole Euclidean spacetime, both quantities differ by singular terms \cite{LEO},
\begin{equation}
H_{\mu \nu}=\partial_\mu C^{(n)}_\nu -\partial_\nu C^{(n)}_\mu
+D_{\mu \nu}
\makebox[.5in]{,}
D_{\mu \nu}=\frac{1}{g} \hat{n}_1. [\partial_\mu,\partial_\nu]\hat{n}_2.
\label{HCD}
\end{equation}
To study magnetic defects, it will also be convenient to consider the associated dual expressions, defining the dual tensors using lower-case letters. For instance, the dual form of the first equation in (\ref{HCD}) reads,
\begin{equation}
h_{\mu \nu}=\epsilon_{\mu \nu \rho \sigma}\partial_\rho C^{(n)}_\sigma+d_{\mu \nu}
\makebox[.3in]{,}
h_{\mu \nu}=\frac{1}{2} \epsilon_{\mu \nu \rho \sigma} H_{\rho \sigma}
\makebox[.3in]{,}
d_{\mu \nu}=\frac{1}{2} \epsilon_{\mu \nu \rho \sigma} D_{\rho \sigma}.
\label{hcd}
\end{equation}

The monopole configurations are obtained from nontrivial $\hat{n}$ mappings \cite{cho-a}, \cite{cho2}-\cite{Shaba},
\begin{equation}
g_m=\oint ds_i\, h_{0i} = \pm \frac{4\pi}{g},
\label{m-ch}
\end{equation}
where the integral is on a surface enclosing a monopole (resp. anti-monopole). The factor of two, with respect to the magnetic charge of a Dirac monopole, is associated with the nonabelian nature of the fields. 

For mappings like these, the term $\vec{L}_{\mu \nu}$ must vanish since $\hat{n}$ does not contain defects localized on two-dimensional worldsheets. On the other hand, the local directions $\hat{n}_1$, $\hat{n}_2$ will be necessarily singular on two-dimensional worldsheets, and therefore they give a nontrivial contribution to $d_{\mu \nu}$ of the form,
\begin{eqnarray}
d_{\mu\nu} &=&\frac{4\pi}{g} \int d\sigma_1 d\sigma_2\, 
\left(\frac{\partial x_w^\mu}{\partial \sigma_1}\frac{\partial x_w^\nu}{\partial \sigma_2}-
\frac{\partial x_w^\mu}{\partial \sigma_2}\frac{\partial x_w^\nu}{\partial \sigma_1}\right) \delta^{(4)}(x-x_{w}(\sigma_1,\sigma_2))\nonumber \\
&=&\frac{4\pi}{g} \int d^2 \sigma_{\mu \nu}\, \delta^{(4)}(x-x_{w}(\sigma_1,\sigma_2)),
\end{eqnarray}
where $x_{w}(\sigma_1,\sigma_2)$ is the Dirac worldsheet.

It will be useful to know that for a monopole/anti-monopole pair localized on the loops ${\cal C}^+$ and ${\cal C}^-$, we have,
\begin{eqnarray}
\partial_\nu d_{\mu \nu}&=&
\frac{4\pi}{g} \left( \oint_{{\cal C}^+} dy_\mu\, \delta^{(4)}(x-y)- \oint_{{\cal C}^-} dy_\mu\, \delta^{(4)}(x-y) \right).
\label{divd}
\end{eqnarray}

\section{Partition functions}
\label{b}

\subsection{Compact $QED(3)$}

The partition function of compact $QED(3)$ with an instanton/anti-instanton pair is,
\begin{equation}
Z = \int  [{\cal D} A] [{\cal D}\Phi] [{\cal D}\bar{\Phi}] F_{gf}\, e^{-S}.
\label{partition_function}
\end{equation}
For example, we can consider the gauge fixing condition,
\begin{equation}
 \partial_\mu (A_\mu +C_\mu) = 0
\end{equation}
introducing a Lagrange multiplier $\beta$, which corresponds to the measure, 
\begin{equation}
F_{gf}=[{\cal D} \beta] e^{i\int d^3x\, \beta\, \partial_\mu (A_\mu + C_\mu)}.
\end{equation}
In order to  single out the terms that depend explicitly on the Dirac string, we linearize the coupling with $C_{\mu}$ by introducing the auxiliary fields $\Lambda_{\mu}$, $\bar{\Lambda}_{\mu}$, and $\lambda_{\mu}$,
\begin{equation}
S = \int d^{3}x \Biggl(\frac{1}{2}\lambda^{2}_{\mu} +
\bar{\Lambda}_{\mu}\Lambda_{\mu} - \frac{i
}{2}\biggl(\bar{\Lambda}_{\mu}D_{\mu}\Phi +
\bar{D}_{\mu}\bar{\Phi}\Lambda_{\mu}\biggr) -
i\lambda_{\mu}\biggl(f_{\mu} +
h_{\mu}\biggr)\Biggr).
\label{11}
\end{equation}

The partition function becomes,
\begin{equation}
Z = \int  [{\cal D} \Psi][{\cal D}\beta] \, e^{ - S_{c}-\int d^{3}x\, \frac{1}{2}\lambda_{\mu}^{2} +
i\int d^{3}x\, (\lambda_{\mu}\,(f_{\mu}+h_{\mu}) - J_{\mu}(A_{\mu}+C_\mu)+\beta\, \partial_\mu (A_\mu +C_\mu))},
\label{partition2}
\end{equation}
where $[{\cal D} \Psi]$ is the measure over all fields, physical and auxiliary, while,
\begin{equation}
S_{c} = \int d^{3}x \biggl(\bar{\Lambda}_{\mu}\Lambda_{\mu} - \frac{i}{2}(\bar{\Lambda}_{\mu}
\partial_{\mu} \Phi +
\Lambda_{\mu}\partial_{\mu}\bar{\Phi})\biggr),
\label{13}
\end{equation}
\begin{equation}
J_{\mu} = \frac{iq}{2}(\bar{\Lambda}_{\mu}\Phi - \bar{\Phi}\Lambda_{\mu}).
\label{14}
\end{equation}
We also note that a constraint is implicit in eq. (\ref{partition2}), because of the $A_\mu$ path integral,
\begin{equation}
\epsilon_{\mu \nu \rho} \partial_\nu \lambda_\rho =J_\mu^c \makebox[.5in]{,}
J_\mu^c = J_\mu +\partial_\mu \beta ,
\label{ampere}
\end{equation}
which implies,
\begin{equation}
\beta =-\frac{1}{\partial^2} \partial_\mu J_\mu.
\label{betaJ}
\end{equation}
That is,
\begin{equation}
Z = \int  [{\cal D} \Psi][{\cal D}\beta] \, e^{ - S_{c}-\int d^{3}x\, \frac{1}{2}\lambda_{\mu}^{2} +
i\int d^{3}x\, [A_\mu (\epsilon_{\mu \nu \rho}\partial_\nu \lambda_\rho-J_\mu^c) + \lambda_{\mu} d_\mu]}.
\label{ZQ-exa}
\end{equation}
where we used,
\begin{eqnarray}
\int d^{3}x\, (\lambda_\mu h_\mu - J_{\mu} C_{\mu}+\beta \partial_\mu C_\mu)=\int d^{3}x\, \lambda_{\mu} (h_\mu -\epsilon_{\mu \nu \rho} \partial_{\nu} C_{\rho}).
\label{17}
\end{eqnarray}

\subsection{$SU(N)$ Yang-Mills in four dimensions}

The Yang-Mills action on the monopole background is,
\begin{eqnarray}
S_{YM}&=&\int d^4x\, \left[ \frac{1}{4} (f_{\mu \nu} +h_{\mu \nu}+k_{\mu \nu})^2 + \frac{1}{2} \bar{g}^{\mu \nu} g^{\mu \nu}\right],
\label{SM}
\end{eqnarray}
where,
\begin{equation}
g^{\mu \nu}=\frac{1}{2} \epsilon_{\mu \nu \rho \sigma}G_{\rho \sigma}=\epsilon_{\mu \nu \rho \sigma}[\partial_\rho+ig(A_\rho+C^{(n)}_\rho)]\Phi_\sigma,~~{\rm etc.}
\end{equation}
Introducing real and complex auxiliary fields, $\lambda_{\mu \nu}$ and $\Lambda_{\mu \nu}$, we obtain,
\begin{eqnarray}
&&S_{YM}=\nonumber \\
&&\phantom{S_{YM}}=S_c+\int d^4x\, \left[\frac{1}{4}\lambda_{\mu \nu} \lambda_{\mu \nu} -\frac{i}{2}
\lambda_{\mu \nu}(f_{\mu \nu}+h_{\mu \nu}+k_{\mu \nu})+i J^\mu (A_\mu+C^{(n)}_\mu) \right],\nonumber \\
&&S_c=\int d^4x\, \left[\frac{1}{2}\bar{\Lambda}^{\mu \nu} \Lambda^{\mu \nu}-\frac{i}{2} (\bar{\Lambda}^{\mu \nu} 
\epsilon^{\mu \nu \rho \sigma}\partial_\rho \Phi_\sigma + {\Lambda}^{\mu \nu} 
\epsilon^{\mu \nu \rho \sigma}\partial_\rho \bar{\Phi}_\sigma)\right],
\end{eqnarray}
that is, the Yang-Mills action with $(A_\mu+C^{(n)}_\mu)$ minimally coupled to the current for charged fields,
\begin{equation}
J^\mu = -\frac{i}{2}\, g \epsilon^{\mu \nu \rho \sigma} \bar{\Lambda}_{\nu \rho}\Phi_\sigma + \frac{i}{2}\, g \epsilon^{\mu \nu \rho \sigma} {\Lambda}_{\nu \rho}\bar{\Phi}_\sigma.
\label{Jlambda}
\end{equation}

\subsection{Gauge fixing}

As gauge fixing, we will adopt the Maximal Abelian gauge (see \cite{verS1} and references therein). For its extension in the context of the Cho-Faddeev-Niemi decomposition, see \cite{kondo6}. Then, for the charged modes we consider,
\begin{equation}
\hat{D}_\mu \vec{X}^{(n)}_\mu =0
\makebox[.5in]{,}
\hat{D}_\mu \vec{X}^{(n)}_\nu=\partial_\mu \vec{X}^{(n)}_\nu+g \hat{A}_\mu\times \vec{X}^{(n)}_\nu ,
\end{equation}
\begin{equation}
\hat{A}_\mu=A_\mu \hat{n}-\frac{1}{g} \hat{n}\times \partial_\mu \hat{n},
\label{Arest}
\end{equation}
while for the diagonal fields, we have,
\begin{equation}
\partial_\mu (A_\mu+C^{(n)}_\mu)=0.
\label{lorentz}
\end{equation}
These conditions can be imposed by means of lagrange multipliers $\vec{b}=b_1 \hat{n}_1+b_2 \hat{n}_2$ and $\beta$, respectively. 

The condition for the charged modes can be rewritten as,
\begin{equation}
{\cal D}_\mu \Phi_\mu =0
\makebox[.5in]{,}
 \bar{{\cal D}}_\mu\bar{\Phi}_\mu =0,
\makebox[.5in]{,}
{\cal D}_\mu=[\partial_\mu +ig(A_\mu+C^{(n)}_\mu)],
\label{MAG}
\end{equation}
so that eqs. (\ref{lorentz}) and (\ref{MAG}) can be implemented by including a factor,
\begin{equation}
e^{i\int_M d^4x\, \left[\beta \partial_\mu (A_\mu+C^{(n)}_\mu)+\bar{b}\, {\cal D}_\mu \Phi_\mu +
b\, \bar{{\cal D}}_\mu \bar{\Phi}_\mu \right]}
\makebox[.5in]{,}
b=\frac{1}{\sqrt{2}}(b_1 +i b_2).
\end{equation} 
We will also have a Faddeev-Popov determinant, exponentiated by means of the associated ghost fields $\vec{c}=c_1 \hat{n}_1+c_2 \hat{n}_2$.  The action for the ghosts contains a term quadratic in $\hat{D}_\mu$,
which can be linearized by considering additional auxiliary fields $\vec{a}^\mu=a^\mu_1 \hat{n}_1+a^\mu_2 \hat{n}_2$. Here, we can also define charged fields, 
\begin{equation}
c=\frac{1}{\sqrt{2}}(c_1 +i c_2)  
\makebox[.5in]{,}
a_\mu=\frac{1}{\sqrt{2}}(a_\mu^1 +i a_\mu^2),
\end{equation}
and introduce a factor whose exponent contains ${\cal D}_\mu$ derivatives linearly (see ref. \cite{kondo6,LEO}).

The final form for the integration measure fixing the above mentioned gauge conditions depends on the combination $A_\mu+C^{(n)}_\mu$, and can be written as,
\begin{equation}
F_{gf}=\tilde{F}_{gf}\,  e^{-i\int d^4x\, (A_\mu+C^{(n)}_\mu)K_\mu} ,
\label{int-measure}
\end{equation}
\begin{equation}
K_\mu=\partial_\mu \beta + \tilde{K}_\mu ,
\end{equation}
where $\tilde{F}_{gf}$ collects all the other factors, independent of $A_\mu +C_\mu$, and the integration measure for ghosts and auxiliary fields. The part of the current $\tilde{K}_\mu$ depends on the charged fields, $a_\mu$, $b$, $c$ and $\Phi_\mu$, and is invariant under U(1) phase transformations of these fields.

In general, for a given gauge field $A^a_\mu$, $a=1,2,3$, many different local frames $\hat{n}_a$ can be introduced to decompose it. 
In refs. \cite{kondo3,kondo7}, Cho variables have been incorporated by including, in the partition function for Yang-Mills theory, an identity written as an integral over local color directions $\hat{n}$, satisfying $\hat{n}.\hat{n}=1$, and then showing that the Jacobian of the transformation,
\[
\vec{A}_\mu, \hat{n} \rightarrow A_\mu, \Phi_\mu, \bar{\Phi}_\mu , \hat{n} ,
\]
is trivial. 

Then, according to the previous discussions, gauge fields with monopole defects are taken into account by considering local color frames where $\hat{n}$ contains defects concentrated on loops. Necessarily, $\hat{n}_a$, $a=1,2$, will be singular on the associated Dirac worldsheets. 

Therefore, the Yang-Mills partition function can be represented as (see ref. \cite{LEO}),
\begin{eqnarray}
Z_{YM} 
& =& \int [{\cal D}A][{\cal D}\Phi][{\cal D}\bar{\Phi}][{\cal D}\hat{n}] F_{gf}\, e^{-S_{YM}} \nonumber \\
&= &\int [{\cal D} \Psi] \tilde{F}_{gf}\, e^{-S_c-\int d^4x\, \frac{1}{4}\lambda_{\mu \nu} \lambda_{\mu \nu}+i\int d^4x\, [A_\mu (\frac{1}{2}\epsilon_{\mu \nu \rho \sigma} \partial_\nu \lambda_{\rho \sigma}-
J^c_\mu )  +\frac{1}{2}\lambda_{\mu \nu}(d_{\mu \nu}+k_{\mu \nu})]},\nonumber \\
\label{ZYMb}
\end{eqnarray}
\begin{equation}
	J^c_\mu=J^\mu +K^\mu,
\end{equation}
where $[D \Psi]$, besides the integration measure for $A_\mu$, $\Phi_\mu$, $\bar{\Phi}_\mu$ and $\hat{n}$, also integrates over the auxiliary fields $\lambda_{\mu \nu}$ and $\Lambda_{\mu \nu}$. Again, because of the path integration over the diagonal field $A_\mu$, we obtain the implicit constraint,
\begin{equation}
J^c_\mu=\frac{1}{2}\epsilon_{\mu \nu \rho \sigma} \partial_\nu \lambda_{\rho \sigma}.
\label{Jconst4}
\end{equation}

\section{Treatment of Dirac strings and worldsheets}
\label{d}

As is well known, in the formalism of first quantization it is simple to express a physical quantity, such as the probability density for the propagation of a particle, in such a way that the Dirac string is no longer apparent. This comes about as in that case the relevant electric current is concentrated on a closed path formed by the composition of the integration path and a given reference path, joining the fixed initial and final particle positions; thus given a relative phase that only depends on the pierced magnetic flux, as long as Dirac quantization condition is imposed.

On the other hand, in a field theory problem, the possibility of representing physical quantities in a way that does not refer to a Dirac string or worldsheet is nontrivial, since the charge current is distributed on the whole Euclidean spacetime. 

In order to obtain a similar result for quantum field theories with a charged sector, we will proceed in three steps. First, we introduce the Hodge decomposition for $\lambda_\mu$ and $\lambda_{\mu \nu}$ so as to isolate the string or worldsheet dependent terms from gauge invariant objects such as their borders, where the monopoles are located. Next, we verify that Dirac strings and worldsheets can be changed by means of an appropriate change of variables associated with a gauge transformation.  Finally, we show that it is always possible to change to an appropriate set of Dirac strings or worldsheets such that the partition function only depends on the monopole positions.

\subsection{The Hodge decomposition} 

In order to isolate the unphysical terms in a physical quantity such as the partition function, we first note that the Dirac string and worldsheet dependence is contained in (cf. eqs. (\ref{ZQ-exa}) and (\ref{ZYMb})),
\begin{equation}
\int d^{3}x\, \lambda_{\mu} d_\mu
\makebox[.5in]{,}
\int d^4x\, \frac{1}{2}\lambda_{\mu \nu} d_{\mu \nu},
\end{equation}
for compact $QED(3)$ and $SU(2)$ Yang-Mills, respectively.

In the first case, it will be convenient to consider the following decomposition,
\begin{equation}
\lambda_{\mu}=\partial_\mu \phi+ B_{\mu},
\end{equation}
with,
\begin{equation}
\partial_\mu B_{\mu}=0 ,
\label{g-fixing}
\end{equation}
and because of eq. (\ref{ampere}), we also have the implicit constraint,
\begin{equation}
\epsilon_{\mu \nu \rho} \partial_\nu B_\rho =J_\mu^c .
\label{curr3}
\end{equation}
Therefore, we have,
\begin{equation}
\int d^{3}x\, \lambda_{\mu} d_\mu = g_m [(\phi (x^-)-\phi(x^+)] + \int d^{3}x\, B_{\mu} d_\mu ,
\label{3dim}
\end{equation}
\begin{equation}
\int d^{3}x\, B_{\mu} d_\mu = g_m\int_{[x_{s}]} dx_{\mu}\, B_{\mu}.
\label{18}
\end{equation}
We should also change the measure appropriately in (\ref{ZQ-exa}),
\begin{equation}
[{\cal D}\lambda]\to [{\cal D}B][{\cal D}\phi] F^{B}_{gf},
\label{delambda}
\end{equation}
where $F^{B}_{gf}$ is the part of the measure fixing the condition $\partial_\mu B_{\mu}=0$,
\begin{equation}
F^{B}_{gf}=[{\cal D}\xi] e^{i\int d^4x\, \xi \partial_\mu B_{\mu}}.
\label{xi-fixing}
\end{equation}

Similarly, for $SU(2)$ Yang-Mills in four dimensions we decompose the auxiliary field $\lambda_{\mu \nu}$ in the following way,
\begin{equation}
\lambda_{\mu \nu}=\partial_\mu \phi_\nu-\partial_\nu \phi_\mu +B_{\mu \nu},
\end{equation}
\begin{equation}
\partial_\mu \phi_\mu=0  \makebox[.5in]{,}  \partial_\nu B_{\mu \nu}=0,
\label{g-fixing2}
\end{equation}
with the implicit constraint, 
\begin{equation}
\frac{1}{2}\epsilon_{\mu \nu \rho \sigma} \partial_\nu B_{\rho \sigma}=J^c_\mu.
\label{curr4}
\end{equation}
That is, 
\begin{equation}
\int d^4x\, \frac{1}{2}\lambda_{\mu \nu} d_{\mu \nu}=\frac{4\pi}{g} \left( \oint_{{\cal C}^+} dy_\mu\, \phi_\mu- \oint_{{\cal C}^-} dy_\mu\, \phi_\mu \right) +\int d^4x\, \frac{1}{2} B_{\mu \nu} d_{\mu \nu},  
\label{dirac.term}
\end{equation}
\begin{equation}
\int d^4x\, \frac{1}{2} B_{\mu \nu} d_{\mu \nu} = \frac{4\pi}{g} \int_{[x_w]} d^2 \sigma_{\mu \nu}\, B_{\mu \nu} .
\label{184}
\end{equation}

The first term in eq. (\ref{dirac.term}) depends on the (gauge invariant) monopole locations, while the Dirac string and worldsheet have been isolated in the second term.

\section{Getting rid of Dirac strings and worldsheets}
\label{in}

\subsection{Compact $QED(3)$}
 
In compact $QED(3)$, let us consider the change of variables,
\begin{equation}
\Phi' = e^{iq\chi}\,\Phi
\makebox[.5in]{,}
A'_\mu = A_\mu + \chi_\mu,
\label{change-abe}
\end{equation}
which has a trivial Jacobian. The phase $\chi$ is multivalued; when we go along any loop encircling a 
closed Dirac string $\partial \Sigma$, given by the border of a surface $\Sigma$, it changes an amount $\Delta \chi$.

In order for $e^{i q\,\chi}$ be single-valued, we must have,
\begin{equation}
q\Delta \chi =2n \pi.
\label{Dcon}
\end{equation}
Under this condition, $e^{i q\,\chi}$ is continuous on any $\Sigma$, so that we obtain,
\begin{equation}
\partial_\mu e^{i q\,\chi}=i q\,   e^{i q\,\chi}\, \chi_\mu ,
\label{derivada}
\end{equation}
where $\chi_\mu$ is locally given by $\partial_\mu \chi$, containing no $\delta$-distribution localized on $\Sigma$. 

Now, under the change in eq. (\ref{change-abe}), the transformed action is,
\begin{equation}
S' = \int d^{3}x \Biggl(\bar{D}_{\mu}\bar{\Phi}D_{\mu}\Phi +
\frac{1}{2}(f_{\mu} + h'_{\mu})^{2} \Biggr),
\label{action'}
\end{equation}
\begin{equation}
h'_{\mu}=h_{\mu}+\epsilon_{\mu\nu\rho}\partial_{\nu}  \chi_{\rho}.
\label{hh}
\end{equation}
As $\partial_\mu h'_{\mu}=\partial_\mu h_{\mu}$, no new monopoles are introduced in this process. The second term in  eq. (\ref{hh}) only represents a flux concentrated on the closed Dirac string $\partial \Sigma$,
\begin{equation}
\pm g_m=\int dS_{\mu}\, \epsilon_{\mu\nu\rho}\partial_{\nu}\chi_{\rho}  =\oint_l dx_{\mu}\, \chi_{\mu}  =\Delta \chi,
\label{multi}
\end{equation}
where the first integral is done over a surface which is crossed by $\partial \Sigma$, so that its border is a loop $l$ encircling $\partial \Sigma$. In particular, this transformation can be used to change the string attached to monopoles from $d_{\mu}$ to $d'_{\mu}$, by choosing, 
\begin{equation}
\epsilon_{\mu\nu\rho}\partial_{\nu} \chi_{\rho} = d'_{\mu}-d_{\mu}.
\label{8}
\end{equation}
Of course, considering eq. (\ref{Dcon}) and the multivaluadness of $\chi$ in eq. (\ref{multi}), Dirac's quantization condition (\ref{dirac_condition}) is obtained. 

At the quantum level, in the representation of $Z$ (cf. eq. (\ref{ZQ-exa})), we have also introduced a charged field $\Lambda_\mu$. Performing the change of variables given in eq. (\ref{change-abe}), together with, 
\begin{equation}
\Lambda'_\mu = e^{iq\chi}\,\Lambda_\mu ,
\end{equation}
we obtain,
\begin{equation}
Z = \int  [{\cal D} \Psi][{\cal D}\beta] \, e^{ - S_{c}+i\int d^{3}x\, \chi_\mu \, J_\mu
-\int d^{3}x\, \frac{1}{2}\lambda_{\mu}^{2} +
i\int d^{3}x\, [(A_\mu +\chi_\mu )(\epsilon_{\mu \nu \rho}\partial_\nu \lambda_\rho-J_\mu^c) + \lambda_{\mu} d_\mu]}. \label{Zqed}
\end{equation}
Now, according to eq. (\ref{ampere}), the only difference between $J_\mu$ and $J_\mu^c$ is $\partial_\mu \beta$. Therefore, we can replace $J_\mu \rightarrow J_\mu^c$ in the exponent of eq. (\ref{Zqed}), since the difference is
\begin{equation}
\int d^3 x\, \chi_\mu  \partial_\mu \beta = -\int d^3 x\, \beta\, \partial_\mu \chi_\mu, 
\end{equation}
which can be nullified by means of a multivalued phase such that,
\begin{equation}
\partial_\mu \chi_\mu =\partial_\mu \partial_\mu \chi = 0.
\end{equation}
The possibility of such a choice will be discussed in the next subsection.

That is, we obtain,
\begin{equation}
Z = \int  [{\cal D} \Psi][{\cal D}\beta] \, e^{ - S_{c}
-\int d^{3}x\, \frac{1}{2}\lambda_{\mu}^{2} +
i\int d^{3}x\, [(A_\mu +\chi_\mu)\epsilon_{\mu \nu \rho}\partial_\nu \lambda_\rho + \lambda_{\mu} d_\mu]}.
\end{equation}
Finally, integrating by parts the term containing $\chi_\mu \epsilon_{\mu \nu \rho}\partial_\nu \lambda_\rho$, and recalling eq. (\ref{8}), we obtain the partition function in (\ref{ZQ-exa})
where $d_\mu$ is replaced by $d'_\mu$, thus showing the independence of $Z$ on the Dirac string choice.

\subsection{Yang-Mills}
\label{YM-sub}

In $SU(2)$ Yang-Mills, let us consider a gauge transformation of the gauge field $\vec{A}_\mu$ given in eq. (\ref{dec}), decomposed in terms of a general frame $\hat{n}_a$,
\begin{equation}
\vec{A}^S_\mu.\vec{T}= S\vec{A}_\mu.\vec{T}S^{-1}+\frac{i}{g}S\partial_\mu S^{-1},
\end{equation}
which has a trivial Jacobian.

As we have seen in ref. \cite{LEO}, in terms of the Cho-Faddeev-Niemi variables, the gauge transformed field is,
\begin{equation}
\vec{A}^S_\mu=A'_\mu \hat{n}'-\frac{1}{g} \hat{n}'\times \partial_\mu \hat{n}' + X^1_\mu\, \hat{n}'_1+ X^2_\mu\, \hat{n}'_2,
\end{equation}
\begin{equation}
A'_\mu=A_\mu+ C^{(n)}_\mu - C^{(n')}_\mu
\makebox[.5in]{,}
\hat{n}'_a = R(S)\, \hat{n}_a,
\label{nonabe-abe}
\end{equation}
where $C^{(n')}_\mu$ is computed with the transformed basis.

In particular, we can consider a singular gauge transformation $S$ along the direction $\hat{n}$, living in the trivial topological sector of $SU(2)$, representing a frame rotation with phase $\chi$. This phase is multivalued when we go along a loop $l$ linking the closed Dirac worldsheet $\partial \Sigma$ to be introduced, given as the border of a three-volume $\Sigma$. In this case, as $\hat{n}'=\hat{n}$, we still have vanishing $\vec{L}'_{\mu \nu}$ (cf. eq. (\ref{FHK})).

For gauge transformations representing a rotation along the $\hat{n}$-axis, that rotates the basis elements $\hat{n}_1$, $\hat{n}_2$ by an angle $\chi$, $C^{(n)}_\mu - C^{(n')}_\mu$ turns out to be $\chi_\mu $, with $\chi_\mu$ locally given by $\partial_\mu \chi$. Similarly to what happens in the compact $QED(3)$ case, $\chi_\mu$ cannot contain singularities (a $\delta$-distribution) on 
any three-volume $\Sigma$. This comes about as $C^{(n)}_\mu$ and $C^{(n')}_\mu$ depend on derivatives of the local color frame (cf. eq. (\ref{Cmu})), which is single valued along any loop $l$.

The second transformation in eq. (\ref{nonabe-abe}) can be equivalently translated to a phase change $\chi$ of the charged sector, which in the partition function representation of eq. (\ref{ZYMb}) includes not only the fields $\Phi_\mu$, $\bar{\Phi}_\mu$ but also the fields $\Lambda_\mu$, $\bar{\Lambda}_\mu$, the charged ghosts, charged lagrange multipliers and charged auxiliary fields in the gauge fixing measure given in eq. (\ref{int-measure}).

Then, the effect of this transformation on the Yang-Mills action is (see ref. \cite{LEO}),
\begin{eqnarray}
S'_{YM}&=&  \int d^4x\, [\frac{1}{4}(f_{\mu \nu}+h'_{\mu \nu}+k_{\mu \nu})^2 + \frac{1}{2} \bar{g}^{\mu \nu} g^{\mu \nu}],\nonumber \\
\label{YM-linha}
\end{eqnarray}
\begin{equation}
h'_{\mu \nu}=h_{\mu \nu}+\epsilon_{\mu \nu \rho \sigma} \partial_\rho \chi_\sigma  , 
\end{equation}
where the second term is localized on $\partial \Sigma$. In particular, to change the Dirac worldsheet attached to monopoles, we should consider,
\begin{equation}
\epsilon_{\mu \nu \rho \sigma} \partial_\rho \chi_\sigma  = d'_{\mu \nu} -d_{\mu \nu},
\end{equation}
representing a trivial flux $4\pi/g$, concentrated on the composition of the initial and final worldsheets.

Then, after performing the change of variables (\ref{nonabe-abe}), we get,
\begin{eqnarray}
Z_{YM} 
& =& \int [{\cal D}A][{\cal D}\Phi][{\cal D}\bar{\Phi}][{\cal D}\hat{n}] F_{gf}\, e^{-S_{YM}} \nonumber \\
&= &\int [{\cal D} \Psi] \tilde{F}_{gf}\, e^{-S_c+i\int d^{3}x\, \chi_\mu \, (J_\mu+\tilde{K}_\mu)}\times \nonumber \\
&& \times e^{-\int d^4x\, \frac{1}{4}\lambda_{\mu \nu} \lambda_{\mu \nu}+i\int d^4x\, [  (A_\mu +\chi_\mu ) (\frac{1}{2}\epsilon_{\mu \nu \rho \sigma} \partial_\nu \lambda_{\rho \sigma}-
J^c_\mu )  +\frac{1}{2}\lambda_{\mu \nu}(d_{\mu \nu}+k_{\mu \nu})]}.\nonumber \\
\label{YMchi}
\end{eqnarray}
Again, with the choice $\partial_\mu \chi_\mu =\partial_\mu \partial_\mu \chi = 0$, 
we can replace $J_\mu +\tilde{K}_\mu \rightarrow J_\mu + \tilde{K}_\mu +\partial_\mu \beta =J_\mu + K_\mu=J_\mu^c$, and similarly to the $QED(3)$ case, we obtain the partition function in (\ref{ZYMb}) where $d_{\mu \nu}$ is replaced by $d'_{\mu \nu}$, thus showing the independence of $Z_{YM}$ with respect to the change of Dirac worldsheet joining the instanton/anti-instanton defects.\\ 

In order to have an explicit form for $\chi_\mu$, we note that it can be associated with a pair of closed center vortices placed at $\partial \Sigma$.  As $\chi_\mu$ does not contain any $\delta$-distribution localized on $\Sigma$, we can use the results given in refs. \cite{engelhardt1,reinhardt} for closed thin center vortices, taking into account the appropriate factors,
\begin{equation}
\chi_{\mu }  = -g_m
\int_{\Sigma } d^{D-1} \tilde{\sigma }_{\nu }
(\delta_{\mu \nu } \partial^{2} - \partial_{\mu } \partial_{\nu } )
D(x-\bar{x} (\sigma )),
\label{dechi}
\end{equation} 
where the (minimum) magnetic charge $g_m$ is given by $2\pi/e$ or $4\pi/g$, in the abelian or nonabelian case, respectively, and $\bar{x} (\sigma )$ is a parametrization of $\Sigma$, a surface or a three-volume in $D=3,4$, respectively. The integration measure is,
\begin{equation}
d^{D-1} \tilde{\sigma }_{\mu}=\frac{1}{(D-1)!} \epsilon_{\mu \alpha_1 ... \alpha_{D-1}} d^{D-1} \sigma_{\alpha_1...\alpha_{D-1}},
\end{equation}
\begin{equation}
d^{D-1} \sigma_{\alpha_1...\alpha_{D-1}}=\epsilon_{k_1 ... k_{D-1}}\frac{\partial \bar{x}_{\alpha_1}}{\partial \sigma_{k_1}}... \frac{\partial \bar{x}_{\alpha_{D-1}}}{\partial \sigma_{k_{D-1}}}\, d\sigma_1 ... d\sigma_{D-1},
\end{equation}
and $D(x)$ is the Green function for the Laplacian operator.

As shown in refs. \cite{engelhardt1,reinhardt}, using Stokes' theorem, $\chi_{\mu }  $ can be written only in terms of 
$\partial \Sigma$ which corresponds to the manifold where the closed Dirac defects are placed ($\partial \partial \Sigma =0$),
\begin{equation}
\chi_{\mu }   = \frac{4\pi }{g}
\int_{\partial \Sigma } d^{D-2} \tilde{\sigma }_{\mu \kappa }
\partial_{\kappa }^{x} D(x-\bar{x} (\sigma ) ).
\end{equation}
For instance, in three dimensions, if a Dirac string along the $z$-axis is considered, we obtain $\chi_0=0$, $\chi_{i}=-(2/g)\,\epsilon_{ij}\partial_j \ln \rho$, which contains no singularity on any plane whose border is the $z$-axis, and can be locally written as, $\chi_{\mu }=(2/g)\, \partial_\mu \varphi $, where $\varphi$ is the multivalued polar angle, in accordance with our previous discussion.
Note also that in general, because of the index structure in eq. (\ref{dechi}), we have $\partial_\mu \chi_\mu =0$.

Finally, it is interesting to discuss the change of variables we have performed here, after the implementation of the MAG gauge fixing condition, in the light of Gribov ideas. In this respect, we would like to underline that there is an important research program based on the implementation of a properly defined path integral, so as to avoid the so called Gribov copies (see refs. \cite{G,UERJ} and references therein). The path integral restriction is usually done by the inclusion of a Gribov-Zwanzinger term to the Yang-Mills action. In fact, this procedure only erases copies connected to each other by infinitesimal gauge transformations, so that even after it is applied, there is still room for large copies living in the trivial topological sector of the theory \cite{H,UERJ}. These are precisely associated with the change of variables we have performed here,  which is along a gauge transformation that lives in the trivial topological sector, as it includes a frame defect such that $\hat{n}_1$, $\hat{n}_2$ rotate twice when we go around the closed Dirac worldsheet. 
Moreover, as shown in ref. \cite{UERJ}, in the case of the MAG, the Gribov region in SU(2) Euclidean Yang-Mills theories can be seen as a cylinder, bounded in all off-diagonal directions, and unbounded along the diagonal one.
Therefore, our procedure would also work after the implementation of the Gribov restriction, as it only involves
operations on the diagonal direction; namely, the use of the implicit constraint (\ref{Jconst4}), derived from the path integration over the diagonal field $A_\mu$, and diagonal gauge transformations with multivalued phase $\chi$.

In other words, the developments in the following section can be seen as a natural way to fix the freedom associated with large copies when Gribov's scenario is applied to the MAG.

\subsection{Decoupling Dirac strings and worldsheets from the charged sector}

Now, it is desirable to express a physical quantity such as the partition function only in terms of observable properties of the monopoles. In this regard, we will show that the line integral in eq. (\ref{18}) can always be nulified for a given choice of Dirac strings, that is, by considering an appropriate change of variables.

As shown in the previous section, when compact $QED(3)$ and $SU(2)$ Yang-Mills theory are considered in the Lorentz and Maximal Abelian gauge, respectively, and a change of variables associated with a multivalued phase satisfying $\partial_\mu \chi_\mu =\partial_\mu \partial_\mu \chi =0$ is performed, the only change in the integrand of the partition function is the substitution,
\begin{equation}
\int d^D x\, \lambda d \to \int d^D x\, \lambda d' = \int d^D x\, \lambda ( d + \epsilon \partial \chi) ,
\label{Dxl}
\end{equation}
where we have simplified the notation by defining,
\begin{equation}
\lambda (d +\epsilon \partial \chi) =\left\{ \begin{array}{ll}
\lambda_\mu (d_\mu + \epsilon_{\mu \nu \rho}\partial_\nu \chi_\rho ) & {\rm or,} \\
\lambda_{\mu \nu} (d_{\mu \nu}+ \epsilon_{\mu \nu \rho \sigma} \partial_\rho \chi_\sigma ) ,& 
\end{array}\right. 
\end{equation}
in $D=3,4$ dimensions, respectively. On the other hand, in section \S \ref{d}, we have introduced a Hodge decomposition of $d$ in terms of the fields $\phi$, $B_\mu$ or $\phi_{\mu}$, $B_{\mu \nu}$ in three and four dimensions, respectively. As $\epsilon \partial \chi$ introduces a closed Dirac string or worldsheet, the borders in $d'$ are the same as in $d$. That is, the 
terms involving $\phi$ and $\phi_\mu$ in eqs. (\ref{3dim}) and (\ref{dirac.term}) do not change after the above mentioned substitution (they are couplings with the gauge invariant monopole locations). Therefore, the only change in those equations is in the couplings of the Dirac defects with the charged sector,
\begin{equation}
\int d^D x\, B d \to  \int d^D x\, B ( d + \epsilon \partial \chi),
\label{Dxb}
\end{equation}
(recall that $\epsilon \partial B$ represents the charged currents, cf. eqs. (\ref{curr3}) and (\ref{curr4})). 

As already discussed, in compact $QED(3)$ and $SU(2)$ Yang-Mills theory, because of the single-valuedness of $e^{i q\,\chi}$ and the local color frame, in the change of variables for $A_\mu$, the function $\chi_\mu$ cannot contain  singularities on the surface or three-volume $\Sigma$ whose border gives the Dirac string or worldsheet. This means that $\chi_\mu$ can be globally written as,
\begin{equation}
\chi_\mu=\partial_\mu \Theta + R_\mu,
\end{equation}
where $\Theta$ coincides with a given branch of $\chi$ on the Euclidean spacetime minus $\Sigma$, and $R_\mu$ is localized on $\Sigma$. When crossing $\Sigma$, $\Theta$ contains a discontinuity, defining a single-valued function, which jumps back to its initial value when we go around any loop linking the Dirac defect $\partial \Sigma$. Therefore, the calculation of $\partial_\mu \Theta$ contains a $\delta$-distribution on $\Sigma$, and $R_\mu$ must be designed to compensate it, giving a nonsingular $\chi_\mu$. 

In this regard, it is useful to consider the formula obtained in refs. \cite{engelhardt1,reinhardt} to separate the so called thin and ideal center vortices, namely,
\begin{equation}
-\int_{\Sigma } d^{D-1} \tilde{\sigma }_{\mu }\,\delta^{(D)}(x-\bar{x} (\sigma ) ) 
-\int_{\Sigma } d^{D-1} \tilde{\sigma }_{\nu }\,(\delta_{\mu \nu } \partial^{2} - \partial_{\mu } \partial_{\nu } )
D(x-\bar{x} (\sigma ) ) = \partial_\mu \Omega,
\end{equation}
where $\Omega$ is the solid angle (normalized to $1$) subtended by $\Sigma$ when viewed from $x$. This solid angle is
single valued when we go along any loop linking $\partial\Sigma$. In other words, using eq. (\ref{dechi}), we obtain, 
\begin{equation}
\Theta = g_m \Omega
\makebox[.5in]{,}
R_\mu=g_m \int_{\Sigma } d^{D-1} \tilde{\sigma }_{\mu }\,\delta^{(D)}(x-\bar{x} (\sigma )).
\end{equation}
As $\Theta$ is single valued, we have $\epsilon_{\mu \nu \rho} \partial_\nu \partial_\rho \Theta =0$, $\epsilon_{\mu \nu \rho \sigma} \partial_\rho \partial_\sigma \Theta =0$, that is, 
\begin{equation}
\int d^Dx\, [\epsilon \partial \chi] B = \int d^Dx\, [\epsilon \partial R] B,
\end{equation}
for any well behaved $B_\mu(x)$. For example, in $D=3$,
\begin{eqnarray}
\int d^3x\, [\epsilon_{\mu \nu \rho}\partial_{\nu}\chi_\rho ] B_\mu &=& g_m \int d^3x\, \int_{\Sigma } d^{2} \tilde{\sigma }_{\rho}\, \epsilon_{\mu \nu \rho}\partial_{\nu}[\delta^{(3)}(x-\bar{x} (\sigma ))B_\mu(\bar{x} (\sigma ))],\nonumber \\ 
&=&  g_m\int_{\partial\Sigma } dy_{\mu}\, B_\mu(y),
\end{eqnarray}
where we used Stokes' theorem. In this manner, we can explicitly verify that $\chi_\mu$ introduces a closed Dirac string $\partial\Sigma$ (cf. eqs. (\ref{18}) and (\ref{Dxb})),
\begin{equation}
\int_{[x_{s}]} dx_{\mu}\, B_{\mu} \to \int_{[x'_{s}]} dx_{\mu}\, B_{\mu}.
\end{equation}
Following a similar procedure in $D=4$, from eqs. (\ref{184}) and (\ref{Dxb}), the change of variables is equivalent to introduce a closed Dirac worldsheet $\partial \Sigma$,
\begin{equation}
\int_{[x_w]} d^2 \sigma_{\mu \nu}\, B_{\mu \nu} \to \int_{[x'_w]} d^2 \sigma_{\mu \nu}\, B_{\mu \nu}.
\end{equation}

Now, in order to show that is is always possible to decouple the Dirac defects from the charged sector in the integrand of the partition functions, let us first consider a simple situation, in three dimensional spacetime, where the charged fields are such that the monopole and the anti-monopole happen to be placed on a given field line of $B_{\mu}$. The field $B_{\mu}$, which satisfies eqs. (\ref{g-fixing}) and (\ref{curr3}), can be seen as a ``magnetic'' field generated by the charge current $J^c_{\mu}$, so that the
associated field lines must be closed and oriented. Now, as the
monopole and the anti-monopole are at the endpoints of the Dirac
strings, we can consider two strings, $[x_s]$ and
$[x'_s]$, contained on the field line, with tangent
vectors oriented parallel or anti-parallel to $B_{\mu}$,
respectively. That is, when we change from $d_\mu$ to $d'_\mu$ we have,
\begin{equation}
P=\int_{[x_s]} dx_{\mu}\, B_{\mu} > 0
\makebox[.5in]{,} N=\int_{[x'_s]} dx_{\mu}\, B_{\mu} <
0 .
\label{20}
\end{equation}
Then, if the system is defined on ${\cal R}^3$, we can deform
continuously $[x'_s]$ into $[x_s]$, keeping
the endpoints fixed. In this process, the line integral of
$B_{\mu}$ will change continuously from a positive to a negative
value, so that an intermediate string must exist such that it is verified,
\begin{equation}
\int_{[x^0_s]} dx_{\mu}\, B_{\mu} = 0.
\label{21}
\end{equation}

We will present a general proof. We start by defining,
\begin{equation}
I_{[x]}=\left\{ \begin{array}{ll}
\int_{[x_{s}]} dx_{\mu}\, B_{\mu}& {\rm or,} \\
\int_{[x_w]} d^2 \sigma_{\mu \nu}\, B_{\mu \nu} .& 
\end{array}\right. 
\end{equation}

Let us consider a Dirac string (worldsheet) $[x]$ joining the anti-monopole and the
monopole, placed at $x^{-}$ and $x^{+}$ (${\cal C}^{-}$ and ${\cal C}^{+}$).
If $I_{[x]}$ is zero, we are done. If not, we
can assume without loss of generality that it gives a positive result. Now, by considering the above mentioned change of variables, we will gain a term,
\begin{eqnarray}
\lefteqn{I_{[\partial \Sigma]} =\int d^D x\, B \epsilon \partial \chi =\int d^D x\, R \epsilon \partial B}\nonumber \\
&& = g_m \int_{\Sigma } d^{D-1} \tilde{\sigma }_{\mu }\,[\epsilon \partial B]_\mu.
\label{Isig}
\end{eqnarray}
If $\epsilon \partial B \equiv 0$ on the whole Euclidean spacetime, this together with the defining property of $B$, ($\partial_\mu B_\mu =0$, $\partial_\nu B_{\mu \nu} =0$) in eqs. (\ref{g-fixing}), (\ref{g-fixing2}), would imply $B$ identically zero, and the term containing the Dirac string would be trivially zero. Therefore, we can suppose that a region of spacetime exists such that $\epsilon \partial B$ is nonzero. In this case, in order to have a nonzero $I_{[\partial \Sigma]}$, it is sufficient to consider $\Sigma$ as a small disk or three-volume placed on that region with $d^{D-1}\tilde{\sigma }_\mu$ oriented along the local direction of $[\epsilon \partial B]_\mu$. Of course, if necessary we can use $-\chi$ instead of $\chi$ so as to render,
\begin{equation}
I_{[\partial \Sigma]} < 0 .
\end{equation}

The important point is that the phase $n \chi$, with $n$ a natural number, also defines a possible singular gauge transformation, as it also leads to a single valued transformation of the charged fields along any closed loop. Therefore, for the associated change of variables, we have,
\begin{equation}
I_{[x']} = I_{[x]} + n I_{[\partial \Sigma]},
\label{25}
\end{equation}
which can be rendered negative for a large enough value of $n$.
Again, $[x']$ can be continuously deformed into
$[x]$, by shrinking $[\partial \Sigma]$ to zero, and in this process
an intermediate string or worldsheet $[x^0]$ must exist such that $I_{[x^0]}=0$ is
verified.\vspace{.1in}

Summarizing, in this section we have shown that it is always possible to make a change of variables with trivial Jacobian, not altering the initial gauge fixing condition, such that the Dirac strings or worldsheets are decoupled from the charged sector of the theory. For instance, in the Cho-Faddeev-Niemi decomposition of $SU(2)$ Yang-Mills theory, this leads to a representation of the partition function where the only effect of Dirac strings is
given by the coupling of the gauge invariant associated borders (monopoles) and the dual field $\phi_\mu$ (see eq. (\ref{dirac.term})), 
\begin{equation}
\int d^4x\, \frac{1}{2}\lambda_{\mu \nu} d_{\mu \nu} \to \frac{4\pi}{g} \left( \oint_{{\cal C}^+} dy_\mu\, \phi_\mu- \oint_{{\cal C}^-} dy_\mu\, \phi_\mu \right).  
\end{equation}
Of course, this procedure simplifies the study of effective monopole ensembles, as discussed in ref. \cite{LEO}. Once the Dirac worldsheets become decoupled, the ensemble integration over the string-like monopoles, can be represented by means of a second quantized complex field $\psi$, coupled to the gauge field $\phi_\mu$ (see refs. \cite{polya1}, \cite{bardakci}-\cite{halpern2}, \cite{antonov} and references therein). In this language, contact interactions between the string-like monopoles, generate a quartic term $\lambda (\bar{\psi}\psi)^2$ which stabilize the system in a phase with spontaneous symmetry breaking, if the correlation between monopoles and the gluon fields generate an effective negative mass term $-m^2 \bar{\psi} \psi$. In the context of the Cho-Faddeev-Niemi decomposition, this effective theory, representing the condensation of monopole degrees of freedom, has been derived in ref. \cite{cho-a} relying on an heuristic treatment of the Dirac worldsheets.

\section{Conclusions}
\label{conc}
Dirac strings and worldsheets are unobservable objects, however, the presence of a charged sector, which in the case of $SU(2)$ Yang-Mills theory is associated with the off-diagonal modes, implies that these unphysical objects appear in the integrand of the partition function representation.

If on the one hand Dirac strings and worldsheets can be changed at will, it would be desirable to have a representation of the partition functions where they are eliminated in favor of their gauge invariant borders, where the monopoles are located.

This is particularly relevant when using the Cho-Fadeev-Niemi gauge field decomposition to guide the obtention of effective theories associated with ensembles of defects. As Dirac worldsheets and center vortices are described as defects of the components $\hat{n}_1$, $\hat{n}_2$ of the local color frame, it is important to have a careful discussion about how to eliminate Dirac worldsheets, and to understand why this procedure cannot be applied to eliminate center vortices. In this respect, note that in effective models constructed only in terms of $\hat{n}=\hat{n}_3$, if on the one hand no information about unphysical Dirac worldsheets is introduced, on the other, the information about the $\hat{n}_1$, $\hat{n}_2$ vortex sector is lost.

In this work, we have seen that Dirac strings in compact $QED(3)$ and Dirac worldsheets in the Cho-Faddeev-Niemi representation of $SU(2)$ Yang-Mills theory, in the Maximal Abelian gauge, can be handled in a similar manner. In particular, the realization of gauge transformations in terms of the Cho-Faddeev-Niemi variables shows that
the consideration of a multivalued phase $\chi$, $\partial_\mu \partial_\mu \chi =0$, has the only effect of including in the integrand of the gauge-fixed partition function a term containing a closed Dirac defect.

In general, by introducing auxiliary fields $B_\mu$, $B_{\mu \nu}$ (representing the char\-ged current $J_\mu^c$) and $\phi$, $\phi_\mu$, for $D=3,4$ respectively, we have been able to isolate the ($B$-dependent) terms where Dirac strings and worldsheets are coupled from those ($\phi$-dependent) where the associated borders (gauge invariant monopole locations) are coupled. 

Then, we have presented the main result of this work, namely, a procedure showing that it is always possible to choose Dirac strings and worldsheets, in such a manner that the $B$-dependent terms vanish. This can be seen as a natural way to fix the remaining freedom, associated with large copies, after the introduction of a Gribov-Zwanzinger term to erase copies connected to each other by infinitesimal gauge transformations. Note that, in the MAG, the Gribov region is a cylinder, bounded in all off-diagonal directions, and unbounded along the diagonal one. Therefore, our procedure also works after the implementation of the Gribov restriction, as it only involves operations on the diagonal direction.

This procedure is specially useful as we are generally interested in studying ensembles of monopoles, so that it is important to write the theory in a form only depending on physical properties of the ensembles to be integrated.
In particular, in the Cho-Faddeev-Niemi decomposition of $SU(2)$ Yang-Mills theory, the ensemble integration, assuming a phase where monopoles condense, is easily related with an effective model for $\phi_\mu$ and a complex field $\psi$ displaying spontaneous symmetry breaking. This model has been obtained in \cite{cho-a}, by following physical heuristic arguments to deal with the Dirac worldsheets, which can be justified by the exact treatment we have presented here to decouple them from the charged sector.

In the presence of a sector of closed center vortices, the $d_{\mu \nu}$ tensor simply gains a term concentrated on the closed thin center vortices \cite{LEO}.  While for the percolating case, in the lattice, closed center vortices display a confining phase exhibiting $N$-ality (see \cite{greensite} and references therein), in the nonpercolating situation they could be associated with Abelian dominance \cite{LEO}.

It is also possible to attach monopoles with a pair of open center vortices carrying flux $2\pi/g$. In the nonpercolating case, center vortex chains would tend to erase magnetic monopoles, forming magnetic dipoles and a nonconfining phase, as occurs in compact $QED(3)$ coupled to massless fermions, where dipoles are formed because of the existence of quasi-zero modes \cite{FO}. On the other hand, from lattice studies \cite{AGG}-\cite{GKPSZ}, the percolating case is a promising phase, possibly displaying not only confinement but also the observed dependence of the confining string tension on the group representation. 

In this regard, it could be argued that the argument in section \S \ref{in} can also be used to get rid of open or closed center vortices, as they would appear in eq. (\ref{Dxl}) parametrized by $d_{\mu \nu}$ (see \cite{LEO}), and an appropriate unobservable closed Dirac worldsheet could be introduced to compensate the center vortex contribution.

However, while for fixed monopole positions it is possible to change the Dirac worldsheet by performing a (singular) topologically trivial $SU(2)$ gauge transformation, for center vortices it is not \cite{LEO}, \cite{engelhardt1,reinhardt}, so that the latter are expected to be physical objects. From the perspective provided by our procedure, this means that a nontrivial correlation between center vortices and charged fields must be generated. This would imply a nontrivial Jacobian for the phase tansformation of the charged fields, precluding the elimination of center vortices by a simple extension of the procedure derived for open Dirac worldsheets. A similar situation applies to the $k_{\mu \nu}$-dependent term in eq. (\ref{YMchi}) (containing nonabelian information): of course, it cannot be eliminated as $k_{\mu \nu}$ depends on the charged fields and the Jacobian for the necessary transformation would be nontrivial. 

 Then, when open or closed physical center vortices are considered, their coupling to the dual field $B_{\mu \nu}$ cannot be made to vanish. In this case, the analysis of the possible phases becomes highly nontrivial, as it involves ensembles of two-dimensional worldsheets correlated with charged fields and loop-like monopoles at the borders. 

\section*{Acknowledgements}
We would like to acknowledge S. P. Sorella and R. Sobreiro for useful discussions. The Conselho Nacional de Desenvolvimento Cient\'{\i}fico e Tecnol\'{o}gico (CNPq-Brazil), the Funda{\c {c}}{\~{a}}o de Amparo {\`{a}} Pesquisa do Estado do Rio de Janeiro (FAPERJ), and the Pr\'o-Reitoria de P\'os-Gradua\c c\~ao e Pesquisa da Universidade Federal Fluminense (PROPP-UFF), are acknowledged for the financial support.


\end{document}